\documentclass[aps,prl,10pt,twocolumn,longbibliography,superscriptaddress]{revtex4-1}

\usepackage{csquotes}
\usepackage{graphicx}
\usepackage{dcolumn}
\usepackage{bm}
\usepackage{hyperref}
\usepackage{gensymb}
\hypersetup{
    colorlinks=true,
    urlcolor= blue,
    citecolor=blue,
linkcolor= blue}

\usepackage{amsmath}
\usepackage{amssymb}

\begin{document}

\title{\textbf{Superconducting magnetoelectric effects in mesoscopic hybrid structures}}

\author{Mostafa Tanhayi Ahari}
\affiliation{Materials Research Laboratory, The Grainger College of Engineering, University of Illinois Urbana-Champaign, Illinois 61801, USA}
\affiliation{Department of Physics and Astronomy $\&$ Bhaumik Institute for Theoretical Physics, University of California, Los Angeles, California 90095, USA}

\author{Yaroslav Tserkovnyak}
\affiliation{Department of Physics and Astronomy $\&$ Bhaumik Institute for Theoretical Physics, University of California, Los Angeles, California 90095, USA}

\begin{abstract}
In superconductors that lack inversion symmetry, a supercurrent flow can lead to nondissipative magnetoelectric effects. We offer a straightforward formalism to obtain a supercurrent-induced magnetization in superconductors with broken inversion symmetry, 
which may have orbital, layer, sublattice, or valley degrees of freedom\textemdash multiband noncentrosymmetric superconductors. The nondissipative 
magnetoelectric effect may find applications in fabricating quantum computation platforms or efficient superconducting spintronic devices. 
We explore how the 
current-induced magnetization 
can be employed to create and manipulate Majorana zero modes in a simple hybrid structure.  
 
\end{abstract}
\maketitle

{\it Introduction}.\textemdash Hybrid structures involving magnetic materials and normal metals are one of the building blocks of spintronic devices. 
Motivated by mitigating dissipation and enhancing device performance, combining superconductivity with spintronics has led to 
interesting equilibrium and nonequilibrium phenomena that fuel interest in superconducting spintronics~\cite{Linder0}. The injection of 
a spin-polarized (super)current into the superconductor, whose transport may be facilitated by an unconventional superconducting order, such as a triplet pairing, has been
of central interest. In this Letter, we primarily focus on the effect of the supercurrent-induced magnetization on an adjacent material in a hybrid structure 
with an emphasis on potential utility in superconducting spintronics and quantum computation devices. 

In multiband superconductors, where electrons have multiple degrees of freedom, such as orbital,
layer, sublattice, or valley, a supercurrent flow may lead to a sizeable induced magnetization.
The source of the induced magnetization is spin polarization~\cite{Edel} 
and/or orbital magnetic
moments~\cite{Niu}, where the former is due to spin-orbit coupling (SOC) and the latter is related to a Berry curvature of the Bloch electronic bands in these materials. Importantly, the supercurrent-induced
orbital magnetization can be orders of magnitude greater than that due to the
spin~\cite{Luca}. Due to the increasing accessibility 
of multiband superconductors with strong SOC, such as TMDs (transition metal dichalcogenides)~\cite{TMD1,*TMD2,*TMD3,*TMD4,*TMD5,*TMD6,*TMD7,*TMD8}, 
or large orbital moments, such as twisted bilayer 
graphene~\cite{Lawtbg1,*Lawtbg2,Watanabe,Law2}, the supercurrent-induced magnetoelectric effects may play a more 
functional role in mesoscopic hybrid structures, particularly in enhancing the efficiency of magnetic memory manipulations~\cite{memory1,*memory2}.

\begin{figure}
  \includegraphics[scale=.065]{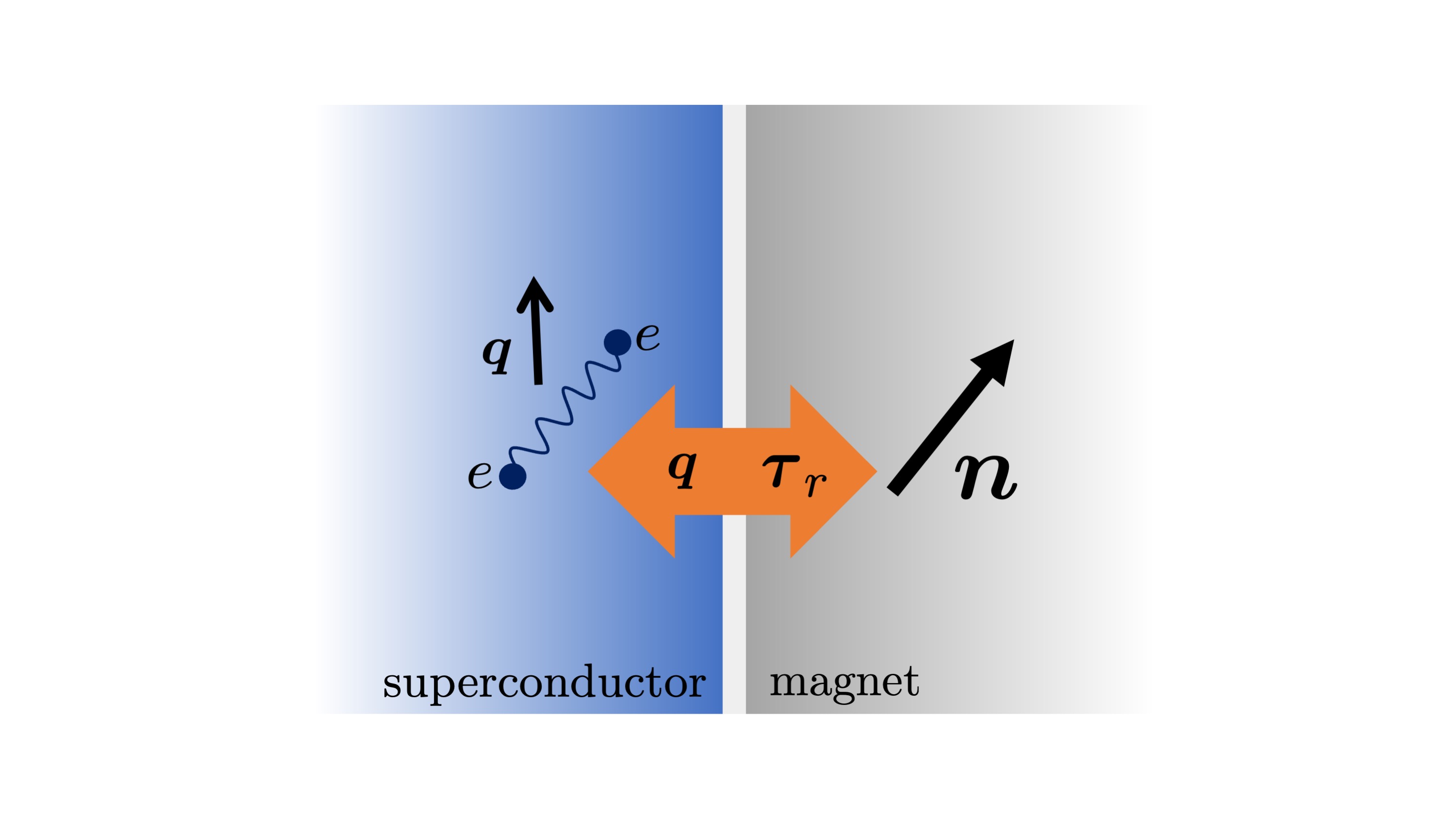}
   \caption{ Schematic of a superconductor-magnet hybrid structure. When the inversion symmetry in the superconductor is broken, 
   a uniform supercurrent with Cooper momentum $2\boldsymbol{q}$ may induce magnetization $\boldsymbol{M}(\boldsymbol{q})$, which can exert a torque $\boldsymbol{\tau}_r\sim \boldsymbol{M}(\boldsymbol{q})\times \boldsymbol{n}$ on the adjacent magnet. Reciprocally, coupling to the magnetization $\boldsymbol{n}$ can induce a supercurrent $\boldsymbol{j}_s$, e.g., for a superconductor with an out-of-plane polar axis $\hat{z}$, one gets $\boldsymbol{j}_s\propto \hat{z}\times \boldsymbol{n}$. 
    } 
   \label{Ons}
\end{figure}

In this Letter, we offer a simple formulation of superconducting magnetoelectric effects for multiband superconductors 
in the static regime, encompassing both orbital and spin magnetization contributions. 
The superconducting magnetoelectric effect has been studied in Ref.~\cite{Law2}, where
the evolution of the magnetoelectric effect across the superconductor-normal metal phase transition is discussed. As we show, our results do not match 
Ref.~\cite{Law2} (in the DC limit) and reveal a greater sensitivity to the characteristics of the pairing gap function. 
Therefore, we suggest that our approach more accurately accounts for the influence of anisotropic pairing gaps. 
Importantly, the magnetization induced by supercurrents in our study may provide additional means to probe the unconventional superconducting pair potentials. Furthermore, we provide an expression for supercurrent flow in multiband superconductors that agrees with earlier results, such as those in Refs.~\cite{Einzel,Qdyn11,Yanaseg}, establishing connections between supercurrent density, Fermi surface parameters, and the superconducting energy gap. 
Additionally, we highlight the reciprocal nature of this phenomenon: Coupling to a dynamic magnetization, such as that of an adjacent magnet, or an applied magnetic field can induce supercurrent. Conversely, supercurrent-induced magnetization can exert torque on the adjacent magnet [see Fig.~\ref{Ons}].

When a superconductor is in contact with a non-superconducting material, a weak superconductivity is induced
in the normal material over mesoscopic distances (superconducting proximity effect)~\cite{Klap}. We will show that the topological character of the 
induced superconductivity in the normal material can be controlled by the magnetoelectric effect in the \enquote{parent}
superconductor. 
The physics here is analogous to the induced topological phase in a quantum wire
discussed in Ref.~\cite{Alicea2}, where zero-energy Majorana bound states are formed in the topological phase. However, here we propose a setup in which the Majorana fermions are electrically created and manipulated via the magnetoelectric effect, without the need for an external magnetic field.
In particular, we propose a simple supercurrent protocol that realizes a braiding operation for the Majorana zero 
modes in a trijunction formed by the wires. Our setup can be relevant for fabricating a platform for quantum 
computation in a web of nanowires~\cite{Alicea3,braid1,*braid2,*braid3,*braid4,*braid5,Beenaker}.

{\it Magnetoelectric effect in multiband superconductors}.\textemdash From the viewpoint of symmetry, a linear 
current-induced magnetization can occur if inversion symmetry is broken~\cite{Law,Moore} 
(a noncentrosymmetric material). As a result, we consider a superconductor with broken inversion symmetry 
described by a time-reversal symmetric normal state Hamiltonian $H_0(\boldsymbol{p})$. 
We assume that close to the Fermi surface, the Bloch band Kramers' splitting due to the broken symmetry  
is much larger than the superconducting gap, 
as is often a realistic limit for the multiband superconductors~\cite{tailor,Sam1}. This assumption justifies
a semiclassical treatment of small perturbations (relative to the band splitting), under which the electron dynamics would be confined within the individual Bloch bands. 
Moreover, in the superconducting state, 
Cooper pairs are formed by time-reversal electron partners belonging to the same Bloch band with opposite momentum $|u_{\nu,\boldsymbol{p}}\rangle$ and $|u_{\nu,-\boldsymbol{p}}\rangle$~\cite{Sig2}, 
where $\nu$ is the band index. It has been shown that the resultant \enquote{intraband} Cooper pairing is the energetically
favorable (stable) superconducting state~\cite{fit,fit2}. As a consequence of projecting physical quantities onto a single band, we focus on the intraband effects, with interband contributions related to quantum metric effects being overlooked~\cite{Qdyn11,Yanaseg,Niu2,Torma,Torma3} (see the Discussion for further details). The Bogoliubov-de Gennes Hamiltonian written in the
Bloch band basis reads
\begin{align}\label{BdG}
\hat{H}(\boldsymbol{p})=
\begin{pmatrix}
    \xi(\boldsymbol{p})&  \Delta(\boldsymbol{p})  \\
     \Delta^*(\boldsymbol{p}) & -\xi(\boldsymbol{p})   
    \end{pmatrix},
\end{align} 
where $\xi(\boldsymbol{p})=\xi(-\boldsymbol{p})$ is a diagonal matrix whose entries are the Bloch 
band energies $\xi_\nu(\boldsymbol{p})$ (measured from chemical potential), which are obtained by a unitary transformation
$U_{\boldsymbol{p}} \, H_0(\boldsymbol{p})\,U^\dagger_{\boldsymbol{p}}=\xi(\boldsymbol{p})$. 
The pairing gap function 
$\Delta(\boldsymbol{p})=U_{\boldsymbol{p}}\,\Delta_0(\boldsymbol{p})U^\text{T}_{-\boldsymbol{p}}$ is a diagonal matrix 
with elements ${\Delta}_\nu(\boldsymbol{p})$, where $\Delta_0(\boldsymbol{p})$ is the usual pairing function written in the space of the electronic degrees of freedom such as 
spin and valley~\cite{supp}. Finally, the quasiparticle energy for band $\nu$ is given by $E_\nu=\sqrt{\xi_\nu^2+|\Delta_\nu|^2}$.   

Under a uniform supercurrent flow, quasiparticle energies get shifted (Doppler shift~\cite{Vol,Tinkham})  as 
$E_\nu+\boldsymbol{q}\cdot \boldsymbol{v}_\nu $, where $\boldsymbol{v}_\nu =\nabla_{\boldsymbol{p}}{\xi}_{\nu}(\boldsymbol{p})$, and $2\boldsymbol{q}$ is the Cooper pair momentum ($\boldsymbol{q}/m_e$ is usually called the superfluid velocity, where $m_e$ is the electron mass). 
It can be checked~\cite{supp} that the total supercurrent and supercurrent-induced magnetization can be written as $\boldsymbol{j}_{s}= \mathbb{T} \cdot \boldsymbol{q}$ and $\boldsymbol{M} = \mathbb{Q}\cdot \boldsymbol{q}$, respectively, where 
\begin{subequations}\label{T}
\begin{align}
    \mathbb{T}_{ij} &= \int d\boldsymbol{\tau}\sum_\nu \Big( \frac{\partial f_\nu}{\partial E_\nu} {v}_{\nu,j}-\frac{\partial n_\nu}{\partial p_j}  \Big) e{v}_{\nu,i} \label{Ta}, \\
    \mathbb{Q}_{ij}&=   \int d\boldsymbol{\tau}\sum_\nu\Big( \frac{\partial f_\nu}{\partial E_\nu} {v}_{\nu,j}-\frac{\partial n_\nu}{\partial p_j}  \Big)  {M}_{\nu,i}.\label{Tb}
\end{align}
\end{subequations}
Here, $f_\nu\equiv f(E_\nu)$ 
is the Fermi-Dirac distribution function which determines the quasiparticle occupancy, $n_{\nu}=\frac{1}{2}\Big(1-\frac{\xi_\nu}{E_\nu}(1-2f_\nu) \Big)$ is the 
occupancy of the electronic single-particle state at band $\nu$ with momentum $\boldsymbol{p}$ in the superconducting state~\cite{deG,Einzel}, 
and $d\boldsymbol{\tau}=d^dp/(2\pi \hbar)^d$ with $d$ being the dimension of the system. $\boldsymbol{M}_{\nu}(\boldsymbol{p})$ is the 
total magnetization pertaining to Bloch band $\nu$ that can be decomposed into spin and orbital~\cite{Moore} components, 
$\boldsymbol{M}_{\nu}(\boldsymbol{p})={\boldsymbol{ s}}_{{\nu}}(\boldsymbol{p})+\boldsymbol{m}_{{\nu}}(\boldsymbol{p})$, where 
\begin{align}\label{spinm}
    \boldsymbol{ s}_{{\nu}}(\boldsymbol{p})=\langle u_{\nu,\boldsymbol{p}}|g\mu_{\text{B}} 
\frac{\boldsymbol{\sigma}}{2}|u_{\nu,\boldsymbol{p}}\rangle
\end{align}
is the spin magnetic moment, and $\boldsymbol{m}_{\nu}(\boldsymbol{p})$ is the orbital
magnetic moment~\cite{Niu1,Pesin} for a 3D crystal 
\begin{align}\label{orbm}
   \boldsymbol{m}_{{\nu}}(\boldsymbol{p})&= \frac{ e\hbar }{2} \text{Im}
   \big \langle \nabla_{\boldsymbol{p}}u_{\nu,\boldsymbol{p}}\big|\times \big[H_0(\boldsymbol{p})-\xi_\nu(\boldsymbol{p})\big]\big|\nabla_{\boldsymbol{p}}u_{\nu,\boldsymbol{p}}\big\rangle.
\end{align}
A nonzero orbital magnetic moment is an intrinsic property of the band that can roughly be interpreted as
a self-rotation of the electron wave function 
around its center of mass~\cite{Niu1}, which could indicate a nonzero 
Berry curvature $\boldsymbol{\Omega}_{\nu,\boldsymbol{p}}=i\hbar^2\langle\nabla_{\boldsymbol{p}}u_{\nu,\boldsymbol{p}}|\times |\nabla_{\boldsymbol{p}}u_{\nu,\boldsymbol{p}}\rangle$ in the band structure.

The terms proportional to $\partial n_\nu/\partial p_j$ in Eqs.~\eqref{T} differ from those obtained
in Ref.~\cite{Law2} (in the DC limit), 
wherein the normal state occupancy of the 
single-particle state at band $\nu$, denoted as $f(\xi_\nu)$, is employed instead of $n_\nu$~\cite{approx}. 
However, Eq.~\eqref{Ta} concurs with the conventional (intraband) superfluid density as established in prior studies~\cite{Einzel,Qdyn11,Yanaseg}, where the terms proportional to $\partial f_\nu/\partial E_\nu$ and $\partial n_\nu/\partial p_j$ are identified, respectively, as the paramagnetic and diamagnetic contributions to the supercurrent~\cite{Tinkham}. This alignment offers reassurance regarding the validity of Eq.~\eqref{Tb} in capturing the intraband current-induced magnetization. 
The dependence of $n_\nu$ on the superconducting pair potential enhances 
the sensitivity of our results to nonuniform (in momentum space) gap functions. 
Given the growing discovery of superconductivity in layered Van der 
Waals materials with anisotropic pairing gap functions, our findings are particularly relevant for understanding how such unconventional gaps influence the magnetoelectric response.

{\it Application to two and four-band models}.\textemdash Let us first consider a two-band system described by a normal state Hamiltonian 
$H_0(\boldsymbol{p})=\xi_{\boldsymbol{p}}\sigma_0+{\bf g}_{\boldsymbol{p}}\cdot \boldsymbol{\sigma}$, 
where $\boldsymbol{\sigma}$ is the vector of Pauli matrices acting on the spin basis, and $\sigma_0$ is a $2\times 2$ identity matrix. As a result of broken inversion symmetry, ${\bf g}_{\boldsymbol{p}}=-{\bf g}_{-\boldsymbol{p}}$, the spin degeneracy of electrons is lifted suggesting
a two-band description of the Fermi surface $\xi_\nu (\boldsymbol{p})=\xi_{\boldsymbol{p}}+\nu |{\bf g}_{\boldsymbol{p}}|$, 
with $\nu=\pm1$ labeling the bands. Finally, it can be checked that the spin magnetic moment of band $\nu$ is given by $ {\boldsymbol{ s}}_{{\nu}}(\boldsymbol{p})=\langle u_{\nu,\boldsymbol{p}}|g\mu_{\text{B}}   \frac{\boldsymbol{\sigma}}{2}|u_{\nu,\boldsymbol{p}}\rangle=\frac{\nu}{2}g\mu_{\text{B}}  \hat{{\bf g}}_{{\boldsymbol{p}}}$.

$H_0(\boldsymbol{p})$ can also describe a minimal model for novel quantum systems such as Weyl
semimetals, in which $\boldsymbol{\sigma}$ acts on orbital space (valence and conduction orbitals mixed with spin), e.g., see~\cite{Burkov,Shen}. In this case, the orbital magnetic moment is given by ${m}_{{\nu},i}(\boldsymbol{p})=-\frac{\hbar e}{2}\varepsilon_{ijl}\, \frac{1}{|{\bf g}_{\boldsymbol{p}}|^2}{\bf g}_{\boldsymbol{p}}\cdot (\partial_{p_j}{\bf g}_{\boldsymbol{p}}\times \partial_{p_l}{\bf g}_{\boldsymbol{p}})$, 
where $\varepsilon_{ijl}$ is the rank-3 Levi-Civita tensor~\cite{Moore}. 

\begin{figure}
  \includegraphics[scale=.08]{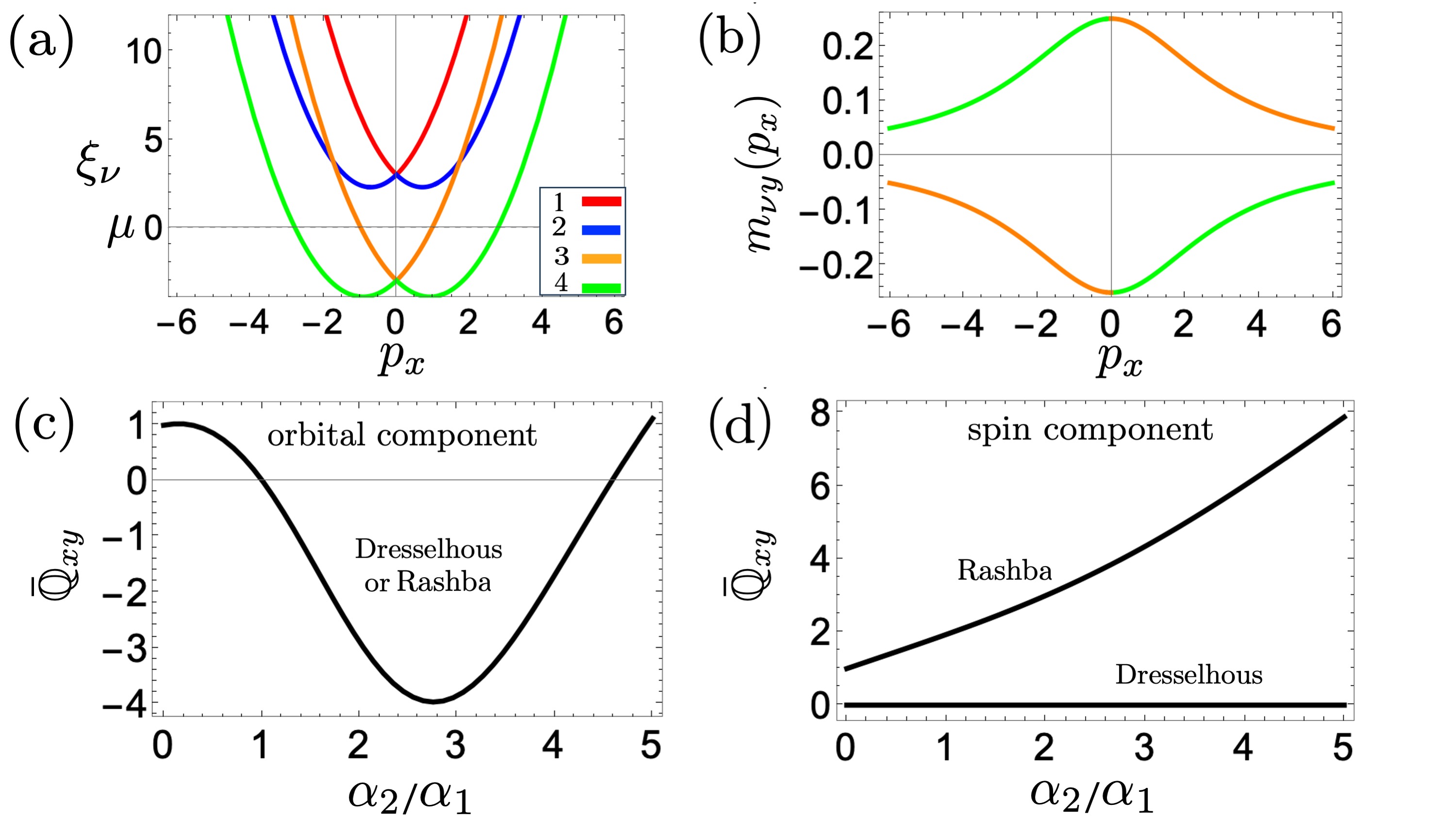}
   \caption{ (a) Schematic of band dispersion in the normal state. The inset shows $\nu=1,2,3$ and $4$ bands, which represent indices $(a_1,a_2)=(1,1),\,(-1,1),\,(1,-1),$ and $(-1,-1)$, respectively. (b) The orbital magnetic moment in the $y$-direction vs $p_x$ for bands $\nu=3,4$. In this plot, we set $\alpha_1=e=t=1$ and $\alpha_2=3$. (c) Orbital component of $\bar{\mathbb{Q}}_{xy}\equiv\mathbb{Q}_{xy}/\mathbb{Q}_{xy}|_{\alpha_2=0}$ vs the strength of SOC ($\alpha_1=1$). The Rashba and Dresselhaus SOC have the same contribution to the orbital magnetization. (d) Spin component of $\bar{\mathbb{Q}}_{xy}$ vs strength of SOC ($\alpha_1=1$). For the Dresselhaus system, the spin component is zero for all $\alpha_2$ values. As a result, to avoid confusion about renormalized $\mathbb{Q}_{xy}$, the plot shows $\mathbb{Q}_{xy}=0$. Here, we use parameters $T=0.1T_c, m_e=0.4, \mu=0, \lambda=3, \Delta=0.5$ (in units of $k_\text{B}T_c$ with $k_\text{B}$ being the Boltzmann constant). 
    } 
    \label{bands}
\end{figure}

Next, we consider a bilayer system with a spin-independent coupling between layers, 
\begin{align}
    H_0=\begin{pmatrix}
        H_1 & V \\
        V & H_2
    \end{pmatrix},
\end{align}
where $H_{i}$ is the Hamiltonian of layer $i=1,2$, and $V=\lambda \sigma_0$ with $\lambda$ being the interlayer interaction~\cite{German}. As illustrative examples, 
we consider each layer to be described by a 2D Rashba or Dresselhaus Hamiltonian $H_{i}=(\frac{p^2}{2m_e}-\mu)\sigma_0+ \bold{g}^i_{\boldsymbol{p}}\cdot \boldsymbol{\sigma}$, where $\bold{g}^i_{\boldsymbol{p}}=\alpha_{i}(p_y,-p_x)$ 
and $\bold{g}^i_{\boldsymbol{p}}=\alpha_{i}(p_x,-p_y)$ for the Rashba and Dresselhaus systems, respectively. 
The band dispersion relation is given by
\begin{align}\label{dispersion}
    \xi_\nu(\boldsymbol{p})=\frac{p^2}{2m_e}-\mu +a_1|\bold{g}^+_{\boldsymbol{p}}|+a_2\sqrt{|\bold{g}^-_{\boldsymbol{p}}|^2+\lambda^2},
\end{align}
where $\bold{g}^\pm_{\boldsymbol{p}}=(\bold{g}^1_{\boldsymbol{p}}\pm \bold{g}^2_{\boldsymbol{p}})/2$ and $a_1,a_2=\pm 1$. Here we identify indices $(a_1,a_2)=(1,1),\,(-1,1),\,(1,-1),$ and $(-1,-1),$ with $\nu=1,2,3$ and $4$, respectively. For the Rashba system, Fig.~\ref{bands}(a) shows the
normal state energy spectrum. 
Eqs.~\eqref{spinm} and \eqref{orbm} can now be used to evaluate the spin and orbital magnetization in the band basis. For the Rashba system~\cite{German}, 
we get
\begin{subequations}
\begin{align}
    \boldsymbol{s}_\nu&=\frac{a_1}{2}g\mu_{\text{B}} \hat{\bold{g}}_{\text{R}} , \\
    \boldsymbol{m}_\nu(\boldsymbol{p})&=\text{sign}[\alpha_+]\frac{a_1}{4}\frac{et \lambda^2\alpha_-}{|\bold{g}^-_{\boldsymbol{p}}|^2+\lambda^2} \hat{\bold{g}}_{\text{R}}\label{Rashbab},
\end{align}
\end{subequations}
where $\hat{\bold{g}}_{\text{R}}=(p_y,-p_x)/p$, $\alpha_\pm=\alpha_1\pm\alpha_2$,
and $t$ is the distance between the layers~\cite{z-direction}. We note that the magnitude of the spin magnetic moment is solely determined by the $g$-factor, while the orbital moment is influenced by multiple parameters, offering greater tunability. Fig.~\ref{bands}(b) shows the
the orbital magnetization in the $y$-direction vs momentum $p_x$ for bands $\nu=3$ and $4$. Note that $\boldsymbol{m}_\nu(\boldsymbol{p})$ 
for bands $\nu=1$ and $2$ is identical to bands $\nu=3$ and $4$, respectively. For the Dresselhaus system, we get
\begin{subequations}
\begin{align}\label{Dress}
    \boldsymbol{s}_\nu&=\frac{a_1}{2}g\mu_{\text{B}} \hat{\bold{g}}_{\text{D}}, \\
    \boldsymbol{m}_\nu(\boldsymbol{p})&=\text{sign}[\alpha_+]\frac{a_1}{4}\frac{et \lambda^2\alpha_-}{|\bold{g}^-_{\boldsymbol{p}}|^2+\lambda^2} \hat{\bold{g}}_{\text{R}} \label{Dressb},
\end{align}
\end{subequations}
where $\hat{\bold{g}}_{\text{D}}=(p_x,-p_y)/p$. It is interesting to note that, unlike the Rashba SOC, in the Dresselhaus SOC, the magnetic moment perpendicular to the current direction is 
only due to the orbital one. Moreover, the orbital moment has the same direction for both systems, as it can 
be seen from Eqs.~\eqref{Dressb} and \eqref{Rashbab}. In both systems, a non-zero orbital magnetic moment arises 
from the fact that the electron state in a band can be a superposition of two layers with different Fermi 
velocities. Consequently, a similar Fermi pocket shift due to Rashba or Dresselhaus SOC can lead to an equivalent orbital magnetic moment. 
To obtain the current-induced magnetization, 
we consider the induced magnetization in the $y$-direction by a current in the $x$-direction. 
Fig.~\ref{bands}(c) and (d) show the orbital and spin component of $\bar{\mathbb{Q}}_{xy}\equiv\mathbb{Q}_{xy}/\mathbb{Q}_{xy}|_{\alpha_2=0}$
vs relative strength of SOC.

{\it Topological superconductivity manipulated by the magnetoelectric effect}.\textemdash Consider a semiconductor wire with Rashba spin-orbit coupling (such as electron-doped InAs~\cite{Rashba}), where proximity to a superconducting substrate induces superconducting pairing. In the presence of an external Zeeman field, the spin degeneracy of electronic bands in the wire is broken, potentially leading to a transition to a topological phase, as discussed in Ref.~\cite{Alicea1}. This transition corresponds to symmetry class D in the Altland-Zirnbauer classification, characterized by a nontrivial $Z_2$ topological number and zero-energy Majorana edge modes.

Additionally, as outlined in Ref.~\cite{Alicea2}, a supercurrent flowing in the substrate superconductor can induce such a transition in the wire. Interestingly, the supercurrent can lower the critical Zeeman field required for the topological phase transition. This suggests the possibility of a topological phase transition solely driven by the supercurrent, without the need for an external Zeeman field, which breaks time-reversal symmetry. Various model superconductors with diverse spin-orbit couplings have been examined in Ref.~\cite{Yanase}, showing that a finite supercurrent alone can trigger a topological phase transition. However, in simple $s$-wave superconductors, a relatively high supercurrent density is necessary for this transition, resulting in the quasiparticle spectrum intersecting the zero-energy level and burying the Majorana states within the bulk continuum.

\begin{figure}
  \includegraphics[scale=.07]{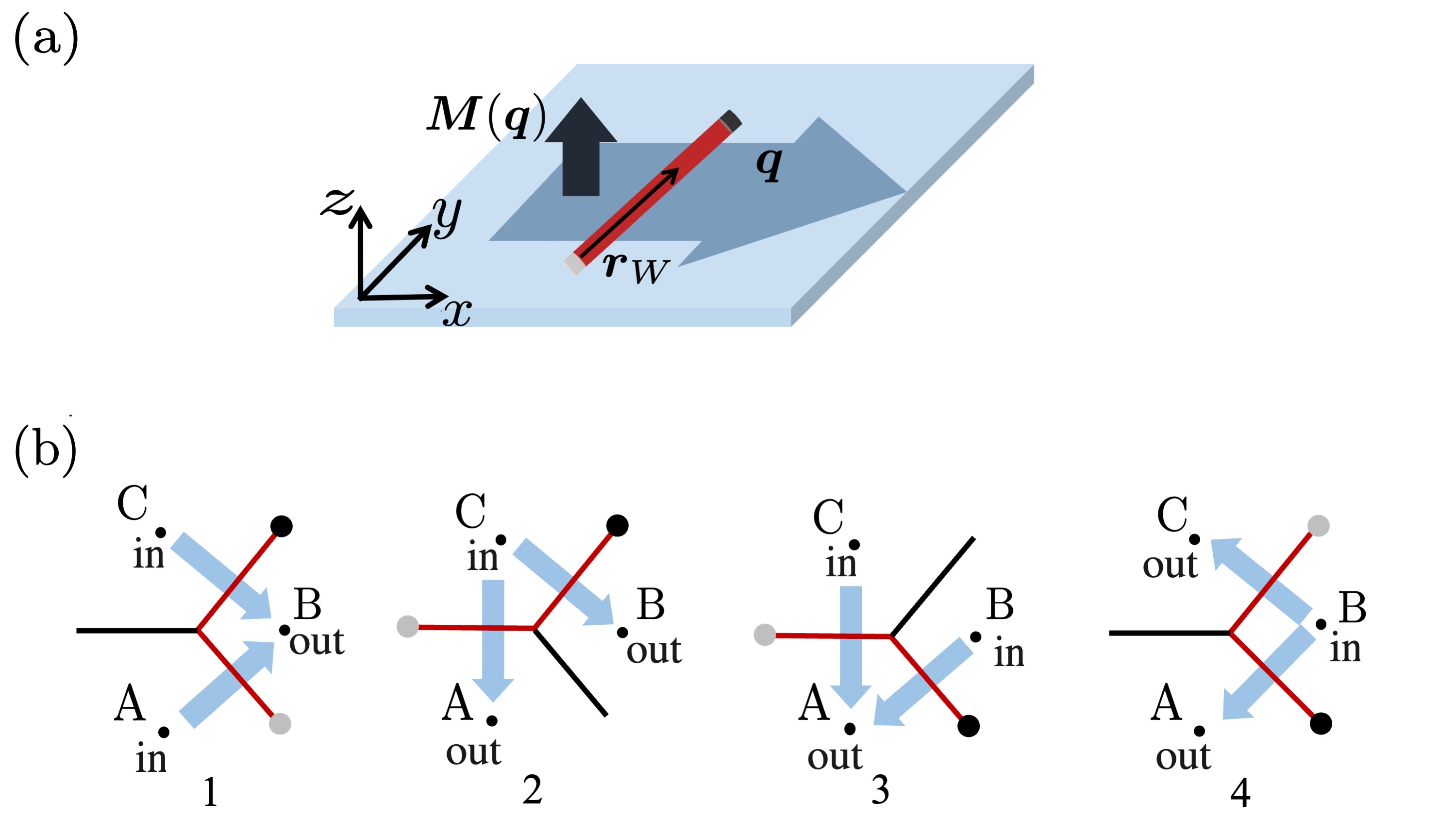}
   \caption{ (a) Schematic of a quantum wire with SOC on a superconductor. The supercurrent flow characterized by $\boldsymbol{q}$ induces a Zeeman field on the wire leading to a topological phase transition. The nontrivial phase has two Majorana modes that appear on both ends of the wire (shown by gray shades). 
(b) Adiabatic exchange of two Majorana bound states. The segments in red are topologically nontrivial ($q>q_T$), where the localized Majorana modes are sketched as the (light and dark shade) grey circles at the ends of the wires. The supercurrent direction and magnitude can be controlled by electrodes connected to the superconductor at points A, B, and C. Starting from a supercurrent between A-B and C-B (configuration 1),
we can decrease the A-B current and simultaneously increase C-A current to transition to configuration 2 adiabatically. Now, decreasing C-B and increasing B-A 
currents results in configuration 3. Similarly, by decreasing C-A and increasing B-C, we return to configuration 1 with the reversed current directions, in which the Majorana modes are exchanged. Repeating the same steps resets the currents and 
leads to a double exchange (braiding) of Majorana modes.
    } 
   \label{tri}
\end{figure}

We propose an alternative approach to topological superconductivity induced and manipulated by 
supercurrent-induced magnetization. We consider a semiconducting wire with Rashba SOC placed on top of a 
superconducting TMD, as shown in Fig.~\ref{tri}(a). Quasi-2D superconducting TMDs with $C_1$, $C_2$, $C_{1v}$, or $C_{2v}$ point group symmetry possess an in-plane polar axis, which allows an in-plane supercurrent to induce an out-of-plane magnetization $\boldsymbol{M}(\boldsymbol{q})$~\cite{Law}. 
The substrate superconductor affects the wire band structure in two ways: an effective Zeeman field on the wire $\sim  \boldsymbol{M}({\boldsymbol{q}})$ and proximity
induced superconductivity with a pairing $\Delta(\boldsymbol{q})=\Delta_0\, e^{i2q_{||} r_W/\hbar}$, 
where $q_{||}$ is the Cooper pair momentum component parallel to the wire and $r_W$ is the position vector along the wire, see Fig.~\ref{tri}(a). 
The extra phase factor leads to a Doppler shift in the quasiparticle spectrum of the wire. For simplicity focusing on $\mu=0$, as discussed in Ref.~\cite{Alicea2}, the boundary between topologically trivial and nontrivial (gapped) phases in a 1D superconductor is when the Zeeman energy in the wire is $E_Z=\sqrt{\Delta_0^2-\alpha_W^2q_{||}^2}>\Delta_0^2/4E_{SOC}$, where $E_{SOC}=m_e\alpha_W^2/2\gg \Delta_0$ is the SOC energy scale in the wire and $\alpha_W$ determines the Rashba SOC in the wire. The lower limit for the Zeeman energy, $\Delta_0^2/4E_{SOC}$, ensures that the phase transition takes place between two trivial and nontrivial gapped phases, which is essential in realizing the isolated Majorana zero modes. Specifically, considering $\Delta_0=0.1$ meV and $E_{SOC}=0.1$ eV, the induced Zeeman energy in the wire must be $E_Z>10^{-4}$ meV. As a result, the topological phase transition may occur at some momentum, denoted by $|\boldsymbol{q}|=q_T$. To enhance the effective Zeeman field on the wire and to maintain $q_T$ below the critical depairing momentum in the substrate superconductor, a semiconductor wire with a large $g$-factor can be employed~\cite{g-factor1,*g-factor2}. 

The supercurrent-induced topological phase may enable the implementation of quantum computation operations using Majorana modes. 
To see this, consider a trijunction fabricated from the wires on the substrate superconductor~\cite{supp}. When a finite supercurrent 
$|\boldsymbol{q}|>q_T$ is crossing a branch of wire in the trijunction, the wire is in the nontrivial phase. Consequently, as sketched in Fig.~\ref{tri}(b), 
an adiabatically slow switching of the supercurrent flow between points A, B, and C leads to the exchange of the Majorana zero modes. The supercurrent switching steps here leads to a similar adiabatic exchange of Majorana fermions first proposed in Ref.~\cite{Alicea3}, where tunable local gate voltages were employed to perform an exchange operation. However, here we propose a supercurrent-induced exchange operation. We note that a double exchange (braiding) operation could realize a $\sigma_z$ gate~\cite{Beenaker}.
If the two Majorana fermions comprising a qubit share no
quasiparticle (even electron parity) the braiding leaves the qubit unchanged. However, if the qubit has an odd electron parity, 
the braiding gives a minus sign. 

{\it Discussion}.\textemdash For a Rashba system with SOC strength of $\alpha_1=2\alpha_2=5\times 10^4$ m/s~\cite{normal}, the estimated spin and orbital magnetizations are $10^{-2}\xi^{-1} \,\,\mu_B/\text{nm}^2$ and $10^{-5}\xi^{-1} \,\,\mu_B/\text{nm}^2$, respectively, where $\xi$ is the superconducting coherence length in nanometers. For a superconductor with $\xi=90$ nm, the corresponding magnetizations would be $10^{-4}  \mu_B/\text{nm}^2$ for spin and $10^{-7}  \mu_B/\text{nm}^2$ for orbital magnetization. We note that measuring this magnetization is feasible within current experimental capabilities, such as employing single-spin microscopy (NV center microscopy)~\cite{NVcenter} or SQUID magnetometry~\cite{Watanabe}.

In a superconductor-magnet hybrid structure, exchange coupling the superconductor to an adjacent magnet, the 
supercurrent-induced magnetization can exert the so-called {\it reactive} torque~\cite{Tser2,Tokat} on the magnet, 
$\boldsymbol{\tau}_r\sim  \boldsymbol{M}(\boldsymbol{q})\times \boldsymbol{n}$ (even under time-reversal), where $\boldsymbol{n}$ is the 
magnetization. Reciprocally, the exchange interaction can modify the supercurrent near the 
interface~\cite{Buzdin}. 
Note that in addition to the reactive torque, a current carrying normal metal with SOC (broken inversion symmetry)
could in general also exert a {\it dissipative} torque on the adjacent magnet, 
$\boldsymbol{\tau}_d\sim \boldsymbol{n}\times  \boldsymbol{M}(\boldsymbol{q})\times \boldsymbol{n}$ (odd under time-reversal). 
However, the dissipative torque is absent in the superconductor-magnet hybrid structure~\cite{distor}.

In Ref.~\cite{Tser1}, a spintronic Josephson phase qubit based on spin superfluidity and spin Hall phenomena is proposed 
in a metal-magnet hybrid structure. The basic idea is that the qubit state can be manipulated by injecting spin current 
(torque), engendered from an electric current flow in the normal metal with SOC~\cite{Zhang1,*Zhang2}. Motivated by eliminating
the Joule heating, a natural generalization of the proposal 
in Ref.~\cite{Tser1} would be to replace the metal with a superconductor. 
The supercurrent can be used for magnetic qubit manipulations and readout. 

In our treatment, we are projecting physical quantities onto a single band, and in doing so, we are neglecting interband coherent effects linked to the quantum metric of the Bloch wave function, discussed in previous 
works~\cite{Qdyn11,Yanaseg,Niu2,Torma,Torma3}. Consequently, without considering these interband effects, 
under a uniform supercurrent superconducting pair potential remains approximately unchanged and the quasiparticle energies experience a Doppler shift.
However, incorporating interband effects introduces an anomalous term for superfluid density, particularly prominent in flat or nearly flat band superconductors~\cite{BernevigT}. Exploring the impact of these geometric effects on supercurrent-induced 
magnetization using our formalism would be an interesting future step. In particular, investigating supercurrent-induced magnetization
in flat-band superconductors could offer an additional means to explore and probe unconventional superconductivity~\cite{Berneviglast}. 

We close this Letter by noting that while we have studied the supercurrent-induced magnetization in a clean superconductor (where physically 
the superconducting gap is much larger than the disorder scattering rate), 
we expect that our results remain qualitatively valid at a low concentration of impurities. 
Indeed as shown in Ref.~\cite{Edirty}, the magnetoelectric effect 
(spin magnetization), although weakened by impurity scattering, is not destroyed in dirty superconductors where the superconducting gap is
smaller than the disorder scattering rate. 

\begin{acknowledgements}
It is a pleasure to acknowledge discussions with Allan H. MacDonald. This work was supported by the NSF under Grant No. DMR-2049979.
\end{acknowledgements}

\bibliography{References}

\begin{thebibliography}{75}
\bibitem [{app()}]{approx}%
  \BibitemOpen
  \href@noop {} {\bibinfo  {journal} {In Ref.~\cite{Law2}, it is assumed that
  the pairing gap is significantly smaller than the band splitting caused by
  inversion symmetry breaking. Consequently, certain expressions are derived
  using the approximation $E_\nu\approx \xi_\nu$ (where the pairing gap is
  considered negligible), resulting in the appearance of $f(\xi_\nu)$ instead
  of $n_\nu$}\ }\BibitemShut {NoStop}%
\bibitem [{z-d()}]{z-direction}%
  \BibitemOpen
  \href@noop {} {\bibinfo  {journal} {In the present 2D system, the Hamiltonian
  does not contain out-of-plane momentum $p_z$. As a result, to evaluate
  derivative with respect to $p_z$ in Eq.~\eqref{orbm} we use~\cite{Niu1}
  \begin{align} \big \langle u_{\nu',\boldsymbol{p}}\big|
  \nabla_{\boldsymbol{p}}u_{\nu,\boldsymbol{p}}\big\rangle= \frac{ \big \langle
  u_{\nu',\boldsymbol{p}}\big|
  \nabla_{\boldsymbol{p}}H_0\big|u_{\nu,\boldsymbol{p}}\big\rangle
  }{\xi_{\nu'}-\xi_{\nu}},\nonumber \end{align} for the interband Berry
  connection (i.e., $\nu\neq\nu'$) with \begin{align} \big \langle
  u_{\nu,\boldsymbol{p}}\big| \frac{\partial H_0}{ \partial
  p_z}\big|u_{\nu',\boldsymbol{p}}\big\rangle \to -\frac{i}{\hbar} \big \langle
  u_{\nu,\boldsymbol{p}}\big|
  [\hat{\mathfrak{z}},H_0]\big|u_{\nu',\boldsymbol{p}}\big\rangle, \nonumber
  \end{align} where
  $\hat{\mathfrak{z}}=\text{diag}(\frac{t}{2},-\frac{t}{2})\otimes \sigma_0$
  can be considered as the out-of-plane component of the position operator with
  $t$ being the distance between the layers~\cite{German,Bahramy}. In
  particular, after some straightforward algebra, we obtain \begin{align}
  m_{\nu y}(\boldsymbol{p}) =& e\text{Re}\Big[ \big \langle
  u_{\nu,\boldsymbol{p}}\big|\hat{\mathfrak{z}}
  \big|u_{\nu,\boldsymbol{p}}\big\rangle \big \langle
  u_{\nu,\boldsymbol{p}}\big|v_x\big|u_{\nu,\boldsymbol{p}}\big\rangle
  \nonumber\\ &- \big \langle u_{\nu,\boldsymbol{p}}\big|\hat{\mathfrak{z}} v_x
  \big|u_{\nu,\boldsymbol{p}}\big\rangle \Big], \nonumber \end{align} where
  $v_x=\frac{\partial H_0}{ \partial p_x}$}\ }\BibitemShut {NoStop}%
\bibitem [{dis()}]{distor}%
  \BibitemOpen
  \href@noop {} {\bibinfo  {journal} {The dissipative torque is also absent in
  a current-carrying 2D massless Dirac system~\cite{Tser2}, which arises on the
  surfaces of strong 3D topological insulators~\cite{Hassan}}\ }\BibitemShut
  {NoStop}%
\end{thebibliography}
\newpage

\pagebreak
\widetext
\begin{center}
\textbf{ Supplementary Material for \\
{\large Superconducting magnetoelectric effects in mesoscopic hybrid structures}}
\end{center}

\begin{center}
Mostafa Tanhayi Ahari$^{1,2}$, Yaroslav Tserkovnyak$^{2}$\\
$^{1}${\it Materials Research Laboratory, The Grainger College of Engineering, University of Illinois, Urbana-Champaign, IL 
61801, USA }\\
$^{2}${\it Department of Physics and Astronomy $\&$ \it Bhaumik Institute for Theoretical Physics, University of California, Los Angeles, California 90095, USA}\\

\end{center}

\setcounter{equation}{0}
\setcounter{figure}{0}
\setcounter{table}{0}
\setcounter{page}{1}
\makeatletter
\renewcommand{\theequation}{S\arabic{equation}}
\renewcommand{\thefigure}{S\arabic{figure}}

\subsection{Magnetoelectric effects in multiband superconductors: static limit}

Here, employing the Bogoliubov-de Gennes formalism, we offer a straightforward derivation of supercurrent-induced magnetization in multiband superconductors. Additionally, we provide an expression for supercurrent flow in multiband superconductors, aligning with the findings of Ref.\cite{Einzel} and Ref.\cite{Qdyn11} (excluding the quantum metric effects). 

Consider the normal state Hamiltonian $H_0(\boldsymbol{p})$. The Bloch band basis representation of the Hamiltonian is achieved by an unitary transformation as 
\begin{align}
    U(\boldsymbol{p})H_0(\boldsymbol{p})U^\dagger (\boldsymbol{p})\equiv \hat{\xi}(\boldsymbol{p})=\text{diag}[\xi_1(\boldsymbol{p}), \, \xi_2(\boldsymbol{p}),\, \dots ],
\end{align}
where $\xi_\nu(\boldsymbol{p})$ is the single-particle energy of Bloch band $\nu$ (measured from the chemical potential in the system). Similarly, we apply the unitary transformation to the Bogliubov-de Gennes (BdG) Hamiltonian of the corresponding superconductor  
\begin{align}
     \begin{pmatrix}
   U(\boldsymbol{p})&  0 \\
     0 & U^*(-\boldsymbol{p})   
    \end{pmatrix} 
    \begin{pmatrix}
   H_0(\boldsymbol{p})&  \Delta(\boldsymbol{p})  \\
     \Delta^\dagger(\boldsymbol{p}) & -H_0^*(-\boldsymbol{p})   
    \end{pmatrix}
    \begin{pmatrix}
   U^\dagger(\boldsymbol{p})&  0 \\
     0 & U^{*\dagger}(-\boldsymbol{p})   
    \end{pmatrix}   =
\begin{pmatrix}
    \hat{\xi}(\boldsymbol{p})& \hat{ \Delta}(\boldsymbol{p})  \\
     \hat{\Delta}^*(\boldsymbol{p}) & -\hat{\xi}(-\boldsymbol{p})   
    \end{pmatrix},
\end{align}
where $ U(\boldsymbol{p}) \Delta(\boldsymbol{p})U^{*\dagger}(-\boldsymbol{p})\equiv \hat{ \Delta}(\boldsymbol{p}) $. The energetically favorable pairing function in the superconducting phase $\Delta(\boldsymbol{p})$ is obtained~\cite{fit,tailor,fit2} by the {\it fitness} criteria $H_0(\boldsymbol{p})\Delta(\boldsymbol{p})-\Delta(\boldsymbol{p})H^*_0(-\boldsymbol{p})=0$. Assuming that the electrons with opposite momentum are degenerate, $\xi_\nu(\boldsymbol{p})=\xi_\nu(-\boldsymbol{p})$, the fitness criteria becomes $\hat{\xi}(\boldsymbol{p})\hat{ \Delta}(\boldsymbol{p})-\hat{ \Delta}(\boldsymbol{p})\hat{\xi}(\boldsymbol{p})=0$. This, in general, implies that the pair potential is diagonal in the band basis (intraband Cooper pairing), $\hat{ \Delta}(\boldsymbol{p})=\text{diag}[\Delta_1(\boldsymbol{p}), \, \Delta_2(\boldsymbol{p}), \, \dots]$. As a result, interband pairings are ignored here. Within an isolated band, the BdG Hamiltonian for a supercurrent-carrying superconductor reads
\begin{align}
   \hat{H}=\frac{1}{2V}\sum_{\boldsymbol{p},\nu}(c_{\boldsymbol{p}+\boldsymbol{q},\nu}^\dagger \, c_{-\boldsymbol{p}+\boldsymbol{q},\nu})
    \begin{pmatrix}
    \xi_\nu(\boldsymbol{p}+\boldsymbol{q})& \Delta_{\nu,\boldsymbol{q}}(\boldsymbol{p})  \\
     \Delta_{\nu,\boldsymbol{q}}^*(\boldsymbol{p}) & -\xi_\nu(-\boldsymbol{p}+\boldsymbol{q})   
    \end{pmatrix}
    \begin{pmatrix}
    c_{\boldsymbol{p}+\boldsymbol{q},\nu} \\
    c_{-\boldsymbol{p}+\boldsymbol{q},\nu}^\dagger
    \end{pmatrix},
\end{align}
where $2\boldsymbol{q}$ is the Cooper momentum. 
To diagonalize this Hamiltonian, we can now apply the Bogoliubov-Valatin transformation, 
\begin{align}
    \begin{pmatrix}
    c_{\boldsymbol{p}+\boldsymbol{q},\nu} \\
    c_{-\boldsymbol{p}+\boldsymbol{q},\nu}^\dagger
    \end{pmatrix}=
    \begin{pmatrix}
        \mathtt{u}_\nu^* & -\mathtt{v}_\nu\\
        \mathtt{v}^*_\nu & \mathtt{u}_\nu 
    \end{pmatrix}
    \begin{pmatrix}
    \alpha_{\boldsymbol{p},\nu} \\
    \beta_{\boldsymbol{p},\nu}^\dagger
    \end{pmatrix},
\end{align}
where $\alpha_{\boldsymbol{p},\nu}$ and $\beta_{\boldsymbol{p},\nu}$ are quasiparticle annihilation operators, and $|\mathtt{u}_\nu|^2=1-|\mathtt{v}_\nu|^2$ with
\begin{align}
    |\mathtt{v}_\nu|^2=\frac{1}{2}\Big(1-\frac{\big(\xi_\nu(\boldsymbol{p}+\boldsymbol{q})+\xi_\nu(\boldsymbol{p}-\boldsymbol{q})\big)/2}{\sqrt{\frac{1}{4}\big(\xi_\nu(\boldsymbol{p}+\boldsymbol{q})+\xi_\nu(\boldsymbol{p}-\boldsymbol{q})\big)^2+|\Delta_{\nu,\boldsymbol{q}}(\boldsymbol{p})|^2}}\Big).
\end{align}
Consequently, we obtain
\begin{align}
    \hat{H}=\frac{1}{2}\sum_{\boldsymbol{p},\nu} \big(E^+_\nu(\boldsymbol{p},\boldsymbol{q}) \alpha_{\boldsymbol{p},\nu}^\dagger \alpha_{\boldsymbol{p},\nu}+E^-_\nu(\boldsymbol{p},\boldsymbol{q})\beta_{\boldsymbol{p},\nu}^\dagger \beta_{\boldsymbol{p},\nu}\big)+\text{const.}
\end{align}
where the positive energies $E^\pm_\nu(\boldsymbol{p},\boldsymbol{q})>0$ (we assume sufficiently small $\boldsymbol{q}$ so that a stable
Cooper pair state exists~\cite{Akbari1,*Akbari2}) are given by
\begin{align}
    E^{\pm}_\nu(\boldsymbol{p},\boldsymbol{q}) =\pm\frac{1}{2}\big(\xi_\nu(\boldsymbol{p}+\boldsymbol{q})-\xi_\nu(\boldsymbol{p}-\boldsymbol{q})\big)+ \sqrt{\frac{1}{4}\big(\xi_\nu(\boldsymbol{p}+\boldsymbol{q})+\xi_\nu(\boldsymbol{p}-\boldsymbol{q})\big)^2+|\Delta_{\nu,\boldsymbol{q}}(\boldsymbol{p})|^2}.
\end{align}

In the presence of a uniform supercurrent, we express a single-particle operator $\hat{\bold{A}}$ in the Nambu space as 
\begin{align}\label{opb}
   \hat{\bold{A}}=\frac{1}{2V}\sum_{\boldsymbol{p},\nu}(c_{\boldsymbol{p}+\boldsymbol{q},\nu}^\dagger \, c_{-\boldsymbol{p}+\boldsymbol{q},\nu})
    \begin{pmatrix}
        \bold{A}_\nu({\boldsymbol{p}+\boldsymbol{q}} ) & 0 \\
        0 & -\bold{A}^*_\nu({-\boldsymbol{p}+\boldsymbol{q}} )
    \end{pmatrix}
    \begin{pmatrix}
    c_{\boldsymbol{p}+\boldsymbol{q},\nu} \\
    c_{-\boldsymbol{p}+\boldsymbol{q},\nu}^\dagger
    \end{pmatrix},
\end{align}
where $\bold{A}_\nu(\boldsymbol{p})$ is the intraband component of operator $\hat{\bold{A}}$ pertaining to band $\nu$. The examples of $\bold{A}_\nu(\boldsymbol{p} )$ include $\nu \hat{\boldsymbol{g}}_{\boldsymbol{p}}$ for spin, $\nabla_{\boldsymbol{p}}\xi_\nu$ for velocity, and $\boldsymbol{m}_{\nu} (\boldsymbol{p})$ for orbital magnetization. Applying the Bogoliubov transformation to Eq.~\eqref{opb} and taking the thermal average, where $\langle \alpha_{\boldsymbol{p},\nu}^\dagger \alpha_{\boldsymbol{p},\nu} \rangle =f(E_\nu^+)$, $\langle \beta_{\boldsymbol{p},\nu}^\dagger \beta_{\boldsymbol{p},\nu} \rangle =f(E_\nu^-)$, and $\langle \alpha_{\boldsymbol{p},\nu}^\dagger \beta_{\boldsymbol{p},\nu} \rangle =\langle \alpha_{\boldsymbol{p},\nu} \beta_{\boldsymbol{p},\nu} \rangle  =0$, we obtain (assuming $\bold{A}_\nu(\boldsymbol{p} )=\bold{A}^*_\nu(\boldsymbol{p} )$) 
\begin{align}
  \langle \hat{\bold{A}} \rangle =\frac{1}{2V}\sum_{\boldsymbol{p},\nu} \Big[\Big(\mathtt{u}_\nu \bold{A}_\nu({\boldsymbol{p}+\boldsymbol{q}}) \mathtt{u}_\nu^* -\mathtt{v}_\nu \bold{A}_\nu({-\boldsymbol{p}+\boldsymbol{q}}) \mathtt{v}_\nu^*\Big) f(E_\nu^+)+ \Big(\mathtt{v}_\nu^* \bold{A}_\nu({\boldsymbol{p}+\boldsymbol{q}}) \mathtt{v}_\nu -\mathtt{u}_\nu^* \bold{A}_\nu({-\boldsymbol{p}+\boldsymbol{q}}) \mathtt{u}_\nu \Big) (1-f(E_\nu^-))\Big]
\end{align}
For small $\boldsymbol{q}$, we obtain $E^\pm_\nu(\boldsymbol{p},\boldsymbol{q}) \approx E_\nu\pm\boldsymbol{q}\cdot \nabla_{\boldsymbol{p}}\xi_\nu+O(q)^2$, where $E_\nu\equiv E^{\pm}_\nu(\boldsymbol{p},0)$, $\xi_\nu\equiv \xi_\nu(\boldsymbol{p},0)$, and $|\Delta_{\nu,\boldsymbol{q}}(\boldsymbol{p})|^2\approx |\Delta_{\nu}(\boldsymbol{p})|^2+O(q)^2$~\cite{Qdyn11}. Assuming that $\bold{A}_\nu(-\boldsymbol{p} )=-\bold{A}_\nu(\boldsymbol{p} )$ (odd under time-reversal symmetry), Taylor expanding and keeping terms linear in $q$, we present the (thermal averaged) quantity as
\begin{align}\label{spino}
   \langle \hat{\bold{A}} \rangle=\frac{1}{V}\sum_{\boldsymbol{p},\nu}  \Big( \frac{\partial f_\nu}{\partial E_\nu}\boldsymbol{q}\cdot \boldsymbol{v}_\nu-\boldsymbol{q}\cdot \nabla_{\boldsymbol{p}}n_\nu  \Big)\bold{A}_\nu({\boldsymbol{p}}),
\end{align}
where $f_\nu\equiv f(E_\nu)$ and $n_{\nu}=|\mathtt{u}_\nu|^2 f_\nu+|\mathtt{v}_\nu|^2 (1-f_\nu)=\frac{1}{2}\Big(1-\frac{\xi_\nu}{E_\nu}(1-2f_\nu) \Big)$ is the occupancy of the single-particle state at momentum $\boldsymbol{p}$ and band $\nu$ in the superconducting state, where $|\mathtt{u}_\nu|^2$ is the probability that the pair state at momentum $\boldsymbol{p}$ is empty, and $|\mathtt{v}_\nu|^2$ is the probability that it is occupied~\cite{Einzel,deG}. 

Now taking the infinite-volume approximation $\frac{1}{V}\sum_{\boldsymbol{p}} \rightarrow \frac{1}{(2\pi \hbar)^2}\int d^2\boldsymbol{p}$ and substituting $\bold{A}_\nu({\boldsymbol{p}})=\bold{M}_\nu({\boldsymbol{p}})$ and $e\bold{v}_\nu({\boldsymbol{p}})$, we obtain the expressions given in the main text. 
We also observe that when dealing with a momentum-independent pairing potential, such as the examples discussed in the main text, the expression above can be rewritten as follows:
\begin{align}\label{spino}
   \langle \hat{\bold{A}} \rangle=\frac{1}{V}\sum_{\boldsymbol{p},\nu}  \Big( \frac{\partial f_\nu}{\partial E_\nu}-\frac{\partial n_\nu}{\partial \xi_\nu}   \Big)(\boldsymbol{q}\cdot \boldsymbol{v}_\nu)\bold{A}_\nu({\boldsymbol{p}}). 
\end{align}

\end{document}